\documentclass[aip,rsi,reprint,graphicx]{revtex4-1}
\usepackage{graphicx}
\usepackage{amssymb}
\usepackage{amsmath}
\usepackage{soul}
\usepackage[usenames]{color}

\ifx\pdftexversion\undefined
\usepackage[dvips]{hyperref}
\else
\usepackage{hyperref}
\fi
\hypersetup{
  colorlinks = true, linkcolor = blue
}

\begin{document}
	
	\title{Fast atom diffraction inside a molecular beam epitaxy chamber, a rich combination.}
	
\author{M. Debiossac}	
\affiliation{Institut des Sciences Mo\'{e}lculaires d’Orsay (ISMO), CNRS, Univ. Paris-Sud, Universit\'{e} Paris-Saclay, Orsay F-91405, France}
\author{P. Atkinson}
\affiliation{Sorbonne Universit\'{e}s, UPMC Univ Paris 06, CNRS-UMR 7588, Institut des NanoSciences de Paris, F-75005, Paris, France}
\author{A. Zugarramurdi}
\affiliation{Institut des Sciences Mo\'{e}lculaires d’Orsay (ISMO), CNRS, Univ. Paris-Sud, Universit\'{e} Paris-Saclay, Orsay F-91405, France}	
\author{M. Eddrief}
\affiliation{Sorbonne Universit\'{e}s, UPMC Univ Paris 06, CNRS-UMR 7588, Institut des NanoSciences de Paris, F-75005, Paris, France}
\author{F. Finocchi}
\affiliation{Sorbonne Universit\'{e}s, UPMC Univ Paris 06, CNRS-UMR 7588, Institut des NanoSciences de Paris, F-75005, Paris, France}
\author{V. H. Etgens} 
\affiliation{Sorbonne Universit\'{e}s, UPMC Univ Paris 06, CNRS-UMR 7588, Institut des NanoSciences de Paris, F-75005, Paris, France}
\author{A. Momeni}
\affiliation{Institut des Sciences Mo\'{e}lculaires d’Orsay (ISMO), CNRS, Univ. Paris-Sud, Universit\'{e} Paris-Saclay, Orsay F-91405, France}
\author{H. Khemliche}
\affiliation{Institut des Sciences Mo\'{e}lculaires d’Orsay (ISMO), CNRS, Univ. Paris-Sud, Universit\'{e} Paris-Saclay, Orsay F-91405, France}
\author{A.G. Borisov}
\affiliation{Institut des Sciences Mo\'{e}lculaires d’Orsay (ISMO), CNRS, Univ. Paris-Sud, Universit\'{e} Paris-Saclay, Orsay F-91405, France}
\author{P. Roncin}
\affiliation{Institut des Sciences Mo\'{e}lculaires d’Orsay (ISMO), CNRS, Univ. Paris-Sud, Universit\'{e} Paris-Saclay, Orsay F-91405, France}

	\begin{abstract}
Two aspects of the contribution of grazing incidence fast atom diffraction (GIFAD) to molecular beam epitaxy (MBE) are reviewed here: the ability of GIFAD to provide \emph{in-situ} a precise description of the atomic-scale surface topology,
and its ability to follow larger-scale changes in surface roughness during layer-by-layer growth.
Recent experimental and theoretical results obtained for the He atom beam incident
along the highly corrugated $[ 1\bar{1}0 ]$ direction of the $\beta_{2}$(2$\times$4)
reconstructed GaAs(001) surface are summarized and complemented by the measurements
and calculations for the beam incidence along the weakly corrugated [010] direction where a periodicity twice smaller as expected is observed.
The combination of the experiment, quantum scattering matrix calculations,
and semiclassical analysis allows in this case to reveal structural characteristics of the surface.
For the in situ measurements of GIFAD during molecular beam epitaxy of GaAs on GaAs surface
we analyse the change in elastic and inelastic contributions in the scattered beam,
and the variation of the diffraction pattern in polar angle scattering.
This analysis outlines the robustness, the simplicity and the richness of the
GIFAD as a technique to monitor the layer-by-layer epitaxial growth.  
	\end{abstract}
	
	\maketitle
	
	\section{Introduction}
	Molecular beam epitaxy (MBE) has played a major role in the development of modern electronic devices. Its ability to deposit successive layers of high purity crystalline materials with monolayer accuracy is well established. Since its development, MBE growth has been monitored \emph{in-situ} by reflection high energy electron diffraction \cite{Ohtake2008} (RHEED).
	More recently, a new diffraction technique using the same geometry
	as RHEED but with keV atoms instead of electrons has emerged as a
	tool to measure the surface crystalline order on metals \cite{Bundaleski2008,Schuller2009}, semi-conductor \cite{Debiossac2014,Khemliche2009} and insulators \cite{Schuller2007,Rousseau2007}.
	To test this technique under conventional semiconductor growth conditions,
	we have attached a GIFAD setup to a conventional III-V MBE
	chamber \cite{Atkinson2014,Debiossac2014}, and studied the surface reconstructions
	and dynamics of layer by layer growth for the homoepitaxy of GaAs.
	
	The paper is organized as follows, firstly a discussion of the GIFAD technique and underlying theory is given, followed by details of its experimental  implantation on a commercial MBE system. The measurement of the surface corrugation of a complex surface reconstruction: the $\beta_2 ( 2\times 4)$ reconstruction of the GaAs (001) surface is then discussed. Finally an analysis of the change in elastic and inelastic scattering, and the variation in polar angle intensity distribution of the scattered He atoms during layer-by-layer growth is presented.	
    	  
	\section{Grazing incidence fast atom diffraction}
\begin{figure}[ht]	\begin{center} \includegraphics [width=80mm]{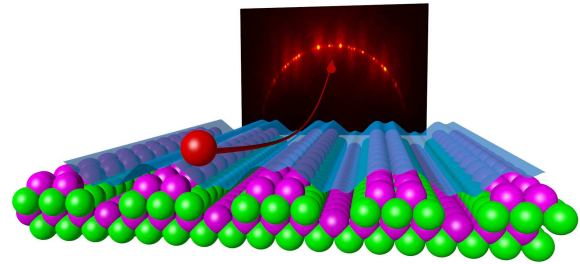} \end{center}
	\caption{An artist view of a fast helium atom diffracting on the rows of well aligned atoms of the $\beta_{2}$(2$\times$4) GaAs surface along the  $[ 1\bar{1}0 ]$ direction.
		The hard corrugated wall model resulting from the projectile-surface potential
		averaged in the direction of the fast motion is displayed here as a blue sheet.
		It is often used to get simple intuitive insight into the GIFAD data.}
	\label{GIFAD}
\end{figure}
Being unable to penetrate the topmost layer, thermal energy helium
atoms are, by nature, perfectly surface sensitive. Grazing incidence
fast atom diffraction (GIFAD) uses helium atoms at energies $E_0$
in the range from some hundreds of eV to some keV impinging at the surface
under grazing incidence angles $\theta\simeq 1^{\circ}$.
The full diffraction pattern can be then recorded with high efficiency
on a position sensitive detector. The grazing incidence conditions correspond to
the very different regimes of the projectile motion parallel and
perpendicular to the surface which typically can be treated separately.
The motion parallel to the surface is fast leading merely to the averaging
of the projectile-surface interaction potential along trajectory.
The diffraction is associated with slow motion perpendicular to the surface
which corresponds to
the projectile energy $E_\perp = E_0 \sin^2\theta $ in the sub-eV range.
Indeed, the corresponding wavelength $\lambda_\perp=2\pi/\sqrt{2 M E_\perp}$
is in the \AA~range of the interatomic distances at the surface ($M$ is the
projectile mass). For further insights on the physics behind GIFAD we
address the reader to refs.~\cite{Rousseau2007,Manson2008,Aigner2008,WinterReview}.

One of the advantages of GIFAD over HAS \cite{Estermann1930,Hulpke1992,Toennies97,Farias2004CPL,Jardine2004}in that diffraction is
preserved even at high substrate temperatures.
In HAS the fraction of coherent scattering $I_c/I_{tot}$ is described by the Debye Waller factor DW, $I_c/I_{tot} = e^{-DW}$ with $DW=-2(\delta k_\perp~u_z)^2$ often forcing  experiments to be performed on surface cooled at liquid nitrogen temperature. In GIFAD, the momentum transfer needed for specular reflection $2\delta k_\perp$ is spread over N successive atoms of the surface so that the effective thermal amplitude is $u_\textrm{eff}=u_z/\sqrt{N}$ and the effective Debye Waller factor is now N times smaller $DW_{eff}=DW/N$ giving rise to a much larger coherence ratio \cite{Rousseau2008,Manson2008}!!.
This explains that both larger perpendicular energies and higher surface
temperatures can be studied in GIFAD. Another specificity of GIFAD is that
only one Laue circle is typically observed\cite{Busch2012} just as if the surface consists
of translation-invariant furrows \cite{Rousseau2007,Zugarramurdi_NIMB_2013,Zugarramurdi2012} (see Fig.\ref{GIFAD}).
Finally, since the interaction reflecting the helium atom from the surface is comparable to that repelling the tip of an AFM, GIFAD can be interpreted in simple topological terms. In this respect it can be metaphorically seen as an AFM operating in the k-space.
	 
	\section{experimental setup MBE and GIFAD}
	
 \begin{figure}[ht]
 	\begin{center}	\includegraphics [width=80mm]{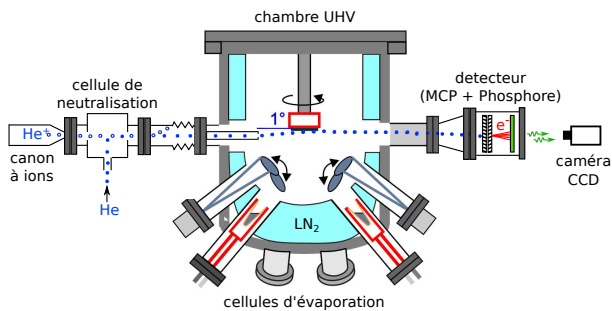}\end{center}
 	\caption{{Schematic view of the MBE chamber, with effusion cells evaporating gallium and arsenic onto a GaAs(001) wafer and the GIFAD setup replacing or complementing RHEED. A beam of He$^+$ ions is extracted at keV from a commercial ion source and then neutralized before entering the chamber. A variable aperture inside the chamber, approx. 10 cm from the sample surface ensures good collimation of the beam. The atoms scattered by the surface are imaged onto a position sensitive detector.}}
 \end{figure} 

The MBE chamber is a standard RIBER Compact21 but a splitting flange has been attached to both the source and detector RHEED ports allowing operation of RHEED or GIFAD independently \cite{Atkinson2014,Debiossac2014}. The GIFAD setup is based on a conventional
VG EX05 hot filament ion source. The $\textrm{He}^+$ ion beam extracted from
the source enters a charge exchange cell filled with helium where $10-20$\% of
the ions are neutralised by resonant electron capture. The ions that survived
the neutralisation are deflected, and the resulting atom
beam is collimated by two sets of diaphragms half a meter apart which also
allow an efficient differential pumping.
These diaphragms might limit the angular divergence of the beam down
to the $0.01^{\circ}$ range, however in price of loss of beam intensity.
The detector is made of one or two microchannel plate electron multipliers facing a phosphor screen which is imaged by a CCD camera. The GIFAD source was attached to the MBE chamber by flexible bellows, and could be "rocked" mechanically using automated motors to vary the incidence angle of the beam on the surface.

Overall the use of GIFAD is comparable to that of RHEED,
with a highly ordered 2D surface giving rise to bright spots
centered on the Laue circle. In GIFAD these spots inherit the
 profile as the incident beam. In addition, the inelastic
background produced by the thermal movement of the surface
atoms\cite{Manson2008,Rousseau2008}, and by the surface
defects \cite{Pfandzelter} gives rise to a low intensity vertical
extensions of the spots as can be seen on Fig.~\ref{GIFAD}.
  		
	\section{Static conditions, high resolution mode}
	 \subsection{The surface electronic density}
	
	 \begin{figure}[ht] 	\begin{center}	\includegraphics [width=80mm]{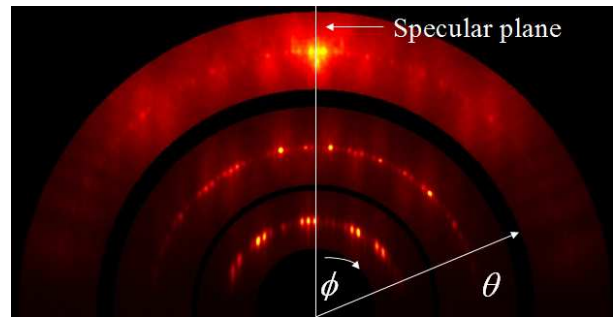}	\end{center}
	 \caption{{Artificial assembly of three diffraction patterns recorded with
	 		400~eV helium atoms incident along the along the $[ 1\bar{1}0 ]$ direction of the $\beta_{2}$(2$\times$4)
	reconstructed GaAs(001) surface at  $530^{\circ}$C. The radii of the different zero-order Laue circles is a direct measure of the polar angle of incidence $\theta$ corresponding to perpendicular energies $E_\perp$ of 17 meV, 55 meV, and 137 meV.}}
		  	\label{fig_raw}
	\end{figure} 
	
When acquisition time is not an issue, the beam divergence can be reduced
down to $0.01^{\circ}$ resulting in a transverse momentum divergence
$\delta k_\perp= k_0 \delta\theta$, and a lateral coherence
$2\pi/\delta k_\perp$ above 20~{\AA}. The drawback is that the beam intensity is also reduced however
extremely rich and complex diffraction patterns with up to almost one hundred
diffraction spots can be recorded with good resolution in a still reasonable
time of few minutes.  	
The unit cell lattice spacing divided by the number of diffraction orders gives a rough measure of the lateral resolution of the surface potential corrugation while a vertical resolution of better than $\lambda_\perp /10$ is easily achieved due to the intrinsic interferometric accuracy.
Under the collimation conditions discussed here, using 400~eV Helium
atoms with incidence angle $< 1^\circ$ yields a  lateral resolution of 0.1~{\AA} and a vertical resolution of 0.01~{\AA}. Such an accuracy challenges the best
theoretical descriptions of the surface, as was demonstrated in a combined
theoretical and experimental GIFAD study of the
$\beta_{2}$(2$\times$4) reconstructed GaAs(001) surface\cite{Debiossac2014}.
In this work, the surface
structure and projectile-surface interaction potential were obtained from
\textit{ab-initio} density functional theory (DFT) calculations. With the DFT inputs,
the diffracted intensities were calculated using a close coupling
technique and compared with experimental data. In the following we shortly review these results.

   \begin{figure}[ht]
   	\begin{center}	\includegraphics [width=80mm]{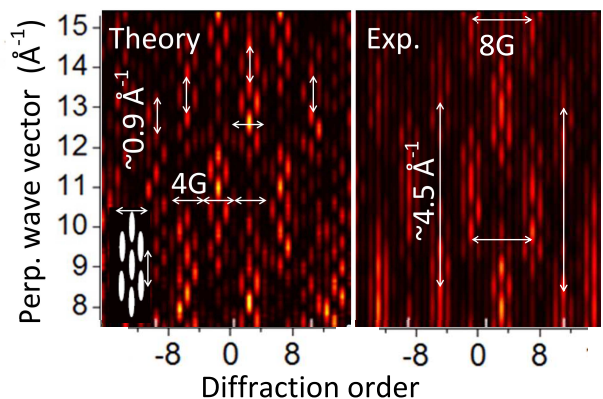}	\end{center}
   	\caption{{Theoretical and experimental quasi specular region of the diffraction charts reporting the evolution of the diffracted intensities with $k_\perp$ (taken from \cite{Debiossac2014}).Both theory and experiment were performed with a misalignment angle corresponding to two reciprocal lattice vector ($|G|=0.39$\AA$^{-1}$). The distinctive chain-like pattern (schematically drawn in the bottom left) composing the experimental motif is well reproduced by the calculations. }}
   	\label{chart_exp}
   \end{figure} 	

In Fig.~\ref{chart_exp} we show a small area of the experimental and 
theoretical diffraction charts corresponding to the $\beta_{2}$(2$\times$4) 
reconstructed GaAs(001) surface. The data shown was collected by varying the 
angle of incidence in the 30--100 meV perpendicular energy range along the
$[1\bar{1}0]$ incidence beam direction. Whereas the repulsive part of the calculated potential 
appeared to bring a good account of the diffraction, the GGA-type 
approximation used for the exchange-correlation potential in the DFT calculations 
cannot correctly describe the long-range Van der Waals interaction. An overall 
good agreement between theory and experiment, as evidenced in Fig.~\ref{chart_exp}, 
could be reached by a simple rescaling of the attractive part of the DFT-derived 
potential. The rescaling resulted in a depth of the physisorption potential well 
of 8.7 meV, which is in line with earlier reports for the same system\cite{Vidali1991}.
This highlights the ability of GIFAD measurements to provide an
strong experimental test of DFT-derived potentials.

\begin{figure}[ht]
		\begin{center}	\includegraphics [width=80mm]{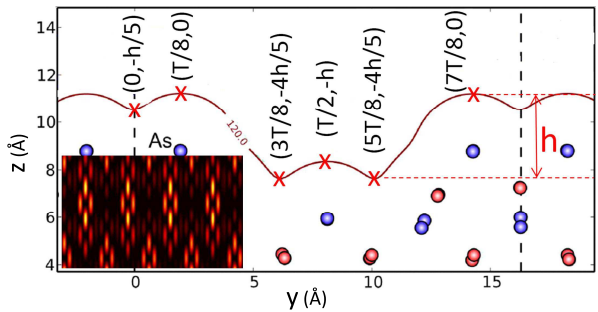}	\end{center}
		\caption{A detailed analysis of the chain-like patterns appearing
			in the central region of diffraction chart in Fig.~\ref{chart_exp}
			indicates that six points are enough to generate such a pattern
			using a ray tracing model. These correspond to the flat sections
			of the corrugation function $z(y)$, the top of the hills and
			bottom of the valleys.}
		\label{6_pts}
	\end{figure}

More interesting is the study of the origin of the chain-like pattern
revealed by the diffraction charts of Fig.\ref{chart_exp}. As shown in
Ref.~(\cite{Debiossac2014}), this pattern can
be interpreted in terms of a simple quasi-classical ray tracing
model and thus reveals the robust information on the structure
of the surface. Let us introduce a reference frame which will be 
used all through the paper in discussion of the diffraction. We 
set $z$-axis perpendicular to the surface and pointing towards the
vacuum, $x$-axis along the axial channel closest to the direction 
of the incident beam, and $y$-axis perpendicular
to this axial channel and lying within the surface plane.
In GIFAD the effective potential ``seen" by the moving projectile
results from the 3D projectile surface interaction averaged
along the fast motion direction $x$. The diffraction results
from the periodicity of the surface potential along $y$-direction
as shown in Fig.~\ref{6_pts} and it is associated with the 
slow motion (with energy $E_\perp$) in 
the $(y,z)$-plane \cite{Rousseau2007,Manson2008,WinterReview}

Now, the quasi specular reflection is governed by the location
of the flat portions of the potential surface, i.e. the top of
the hills and the bottom of the valleys. The diffraction order
$m$ corresponds to the change of the projectile momentum
projection on $y$-axis, $k_y \rightarrow k_y + mG$, where
$G=2\pi/T$ is the reciprocal lattice vector associated with
the period of the structure $T$ in the $y$-direction
(see Fig.~\ref{6_pts}). The period of eight diffraction
orders between two chain patterns in the diffraction chart, indicates
a minimum distance of $T/8$ between diffracting points while
the vertical period of $\delta k_\perp\simeq$ 4.5 \AA$^{-1}$ indicates 
a separation of $\delta z \simeq 0.7$\AA~along $z$ ($\delta k_\perp~2\delta z = 2\pi$). 
The same pattern also appears every four diffraction orders but 
in opposition indicating a contribution of points separated 
by $\delta y =T/4$. The fact that every fourth diffraction order 
is almost dark suggests that these points separated by $T/4$ 
have the same $z$ value so that their contributions add up every $m=4j$ and
cancel exactly every $4j+2$ diffraction orders ($j$ is an
integer) whatever $k_\perp$ is: $m G \delta_y=j\pi/2$.
The same analysis can be performed on the smaller scale details. 
The rapid oscillation with $k_\perp$ every 0.9 \AA$^{-1}$ points to 
a maximum difference of $h=3.5$ \AA~along $z$ between the lowest 
and highest points whereas the quadrature between adjacent orders 
indicates that the top and bottom structures are sitting at $T/2$ of each other. A complete interpretation would have to be more cautious but the careful analysis of the diffraction chart can be quite instructive.

Overall, a semi quantitative presentation can be generated  by a simple six 
point ray tracing model using the topology depicted in Fig.~\ref{6_pts}. 
This interference pattern is extremely 
sensitive to the corrugation amplitudes and, to a lesser extend, 
to the lateral position estimated here around the 0.2 \AA{} range. 
This is made possible via the high redundancy of the diffraction 
chart were the observation of dark lines, nodal structures and 
distinct patterns are robust fingerprints of the symmetry of the 
corrugation function.
Note that, in the present case, the topology was first derived from calculation before it was realized that the observed fingerprint can be interpreted in relatively simple terms. In addition, a full quantum 
diffraction calculation which models the entire diffraction pattern 
is needed to confirm both the surface reconstruction model and 
the helium-surface interaction potential.

\begin{figure}[ht]  \begin{center} \includegraphics [width=70mm]{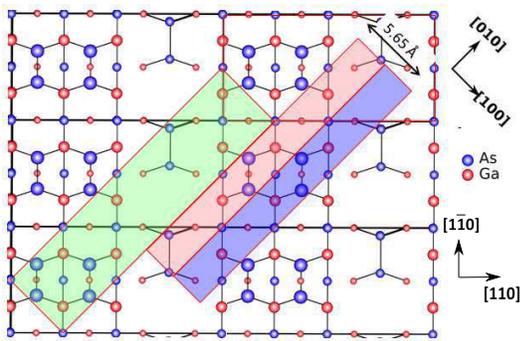}	\end{center}
		\caption{Schematic view of the [010] direction on the $\beta_{2}$(2$\times$4)
			reconstructed GaAs(001) surface. A genuine lattice unit cell is colored in
			green while the two sub unit cells seen as equivalent in GIFAD are colored
			in pink and purple. Both have top atomic plane but
			differ in the lower lying ones.}
		\label{2x4}
	\end{figure}

The $\beta_2 (2\times4)$ reconstruction consists of well aligned pairs
of As dimers separated by a deep valley (see Figs.~\ref{6_pts} and~\ref{2x4})
giving rise to a highly corrugated surface potential along
the $[ 1\bar{1}0 ]$ direction. This is not the case along the [010]
direction oriented at $45^\circ$ from the valleys which are therefore hidden by the dimers of the top layer (see Fig.~\ref{2x4}) producing a much flatter averaged potential in this direction.

\begin{figure}[ht]  \begin{center} \includegraphics [width=80mm]{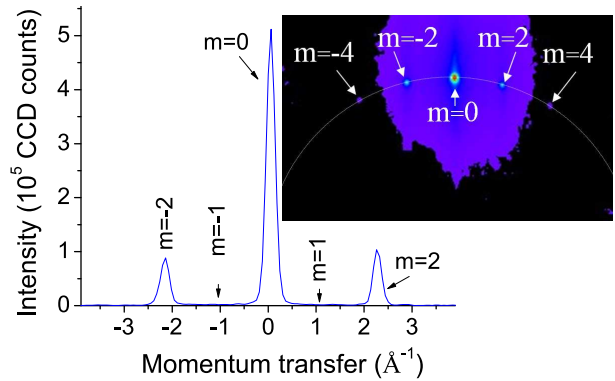} \end{center}
		\caption{Diffraction pattern (inset) recorded with a 400~eV helium atom beam aligned along the $[010]$ direction of  $\beta_{2}$(2$\times$4) reconstructed GaAs(001) surface. The line graph shows to the  elastic intensity profile extracted on the Laue circle \cite{Debiossac2014,Debiossac_V} indicated
			with dotted line in the inset of the figure.The odd diffraction orders are not observed.}
		\label{proj_merged}
	\end{figure}

\begin{figure}[ht]  \begin{center} \includegraphics [width=60mm] {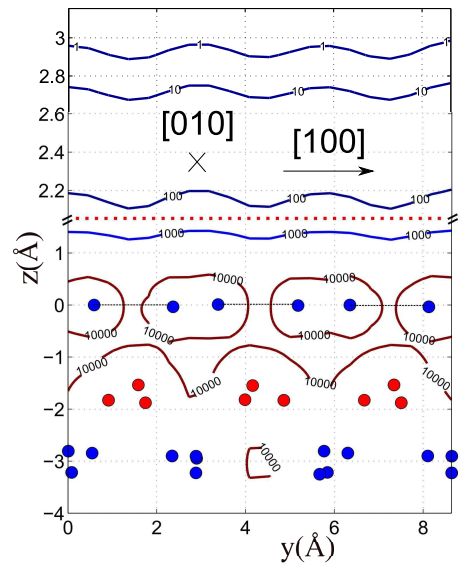}\end{center} 
	\caption{Equipotential lines  $z(y)$ of the 2D potential averaged along the [010] direction:  $V_{2D}(y,z)=E_\perp$ associated with  $E_\perp =1, 10, 100, 1000$ and $10000$~meV. The distance to the surface varies rapidly but the corrugation amplitude is almost constant between 1 and 100 meV. Along this [010] direction, the actual lattice parameter is 5.65 \AA~but top surface layer exhibits a half cell symmetry of 2.83 \AA~corresponding to the As dimers depicted as a dashed line between As atoms (blue).}
	\label{average_pot}
\end{figure}

	The apparent reduction of the surface corrugation when observed along the $[010]$ direction is visible
in Fig.~\ref{proj_merged} where more than $70$\% of the intensity is in the specular spot. The radius of the Laue circle indicates an angle of incidence of $0.55^\circ$ corresponding to a perpendicular energy $E_\perp$ of 38~meV. More surprising, given the reciprocal lattice vector of $G_{[010]}=1.11 \AA^{-1} (2\pi/5.65)$ (Fig\ref{2x4}), only even diffraction orders are observed indicating a half cell pseudo symmetry  $T_{\rm{ps}}=2.82$~{\AA}. The genuine lattice cell is colored in green on Fig.~\ref{2x4} while the half cells observed as equivalent by GIFAD are colored in pink and purple. The pseudo periodicity
	with half-period is also visible in Fig.~\ref{average_pot} which
	displays the interaction potential $V_{2D}(y,z)$ obtained from
	the total projectile-surface interaction potential averaged
	along the fast motion direction $[010]$. Here we use the potentials
	determined in Ref.~\cite{Debiossac2014}.
	This pseudo periodicity can be understood by considering that the deep valleys have a negligible contribution so that the dominant contribution comes from the top As dimers which display an exact half cell period. The bottom of the valley corresponds to the center of the As dimer which is associated with a projected distance of $\simeq1.7${\AA} while the top of the corrugation function corresponds to two As atoms from different dimers which, after projection, are separated by only $\simeq1.17${\AA}.

\begin{figure}[ht]  \begin{center} \includegraphics [width=80mm]{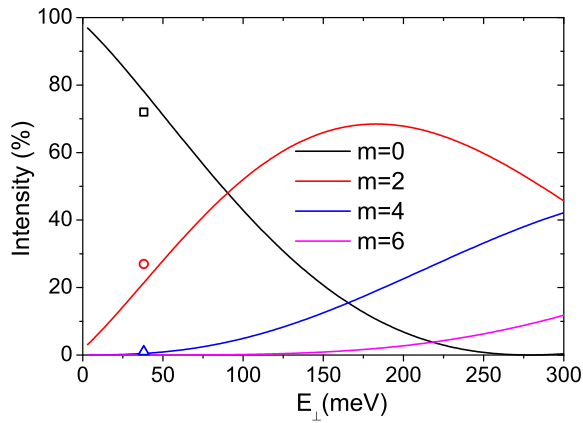}	\end{center}
	\caption{Intensities $I_m$ of the diffraction orders $m$ 
		calculated as function of $E_\perp$ for the 400~eV helium atom 
		beam aligned along the $[010]$ direction of $\beta_{2}$(2$\times$4) 
		reconstructed GaAs(001) surface. The experimental points corresponding 
		to Fig.~\ref{proj_merged} are also reported. Note, that because of the 
		pseudo periodicity (see the text) we compare experimental data obtained 
		for diffraction order $m$ with calculated result $I_{2m}$.}
	\label{AGB}
\end{figure}

	Figure.~\ref{AGB} displays the calculated intensities $I_m$ of the $m=0$ (specular reflection), 
	$m=\pm2,\pm4,\pm6$ diffraction orders for the 400~eV helium atom beam aligned along 
	the $[010]$ direction of  $\beta_{2}$(2$\times$4) reconstructed GaAs(001) surface. 
	For the scattering calculations we used the atomic positions and 
	projectile-surface interaction potential as obtained for the 
	$\beta_{2}$(2$\times$4) reconstruction of GaAs(001) in Ref.~\cite{Debiossac2014}.
	Results are shown as function of the $E_\perp$ energy component. The diffraction 
	order $m$ is defined as the in-plane momentum exchange along $y$-axis (in the 
	direction perpendicular to the incidence direction) given by the reciprocal 
	lattice vector of the native lattice cell 
	$k_y \rightarrow k_y + m~2\pi/(2T_{\rm{ps}})$. Because of the pseudo periodicity
	with half-period, and in full agreement with experimental data, we obtain that 
	the odd diffraction orders are nearly extinct. In addition, for symmetry reasons, 
	$I_m \simeq I_{-m}$. Therefore, without loss of information, for $m \neq 0$ we 
	trace the sum of the intensities of the $\pm m$ diffraction orders. 
	Available experimental data is shown with symbols. Comparison between theory 
	and experiment indicates that the surface corrugation is slightly 
	overestimated in the \textit{ab-initio} calculations, however the overall agreement 
	is quite satisfying.
	
	Along with \textit{ab initio} calculations we have also used the simple hard corrugated wall model (HCW) of Garibaldi \textit{et al} \cite{Garibaldi} well suited for small corrugation. In this model, diffracted intensities are simply derived as a Fourier-like transform of the 1D 
	equipotential line $z(y)$ such that $V_{2D}(y,z(y))=E_\perp$, where	$V_{2D}(y,z)$ is the projectile-surface interaction averaged along the beam direction, here the $[010]$ direction.
	On one hand we have used the HVW to extract a corrugation amplitude from the diffracted intensities via a simple fit, i.e. independently from any a priori surface potential.Taking the simplest sinusoidal approximation $z(y)=z_c~\sin[\frac{2\pi}{T_{\rm{ps}}} y]$, the diffracted 
	intensities $I_m$ are given by the Bessel  functions $I_m=J_m^2(2 k_{\perp} z_c)$. The corrugation amplitude $z_c$ derived from the measured intensity ratio of the diffraction peaks is 
	$0.094$~{\AA} which is in good agreement with the calculated 1D equipotential shown in Fig.~\ref{average_pot}.
	On the other hand, we have also use the HCW to calculate the diffracted intensities from the same potential used in the quantum calculation. The fair agreement allows us to confirm the origin of the observed pseudo symmetry. We have suppressed, one by one, the contributions of the atoms below the As dimers in the construction of the interaction potential to calculate the equipotential lines $z(y)$. The exact locations of these equipotential lines are affected by the underlayers but the diffracted intensities are not. This outline that sensitivity to the top layer has a double origin, not only the contribution from under-layers located $\delta z$ below is weaker by a factor $\simeq e^{-\delta z/R_c}$ ($R_c$ is the typical range of the binary potential) but this contribution tends to be uniform because the spherical contributions converge to a more planar one.  This can be also understood from the surviving 
	Fourier components after averaging the potential: the higher the Miller
	indices of the direction associated to the 2D surface lattice, the higher
	the order of the Fourier components composing the averaged potential. These,
	overall, decay exponentially with increasing order. 
	\cite{Zugarramurdi_NIMB_2013,Celli_1985}

	\section{GIFAD during growth}
	
	This section is devoted to the behaviour of GIFAD during growth and is based on the recent paper by Atkinson \textit{et al.} \cite{Atkinson2014}. One of the most important uses of RHEED for MBE is the calibration of growth rates during layer-by-layer growth using RHEED intensity oscillations. With GIFAD, these oscillations have been investigated in detail for different projectile energies and angles of incidence as well as azimuthal angle and temperature for GaAs homoepitaxy \cite{Atkinson2014}.

	Highly resolved GIFAD images such as the one displayed in Fig.\ref{fig_raw} require over a minute exposure time to achieve a good signal-to-noise ratio mainly due to limitations in the primary beam intensity. The growth oscillations discussed here were however recorded with the beam along the$[110]$ direction, where the apparent corrugation is weaker and the diffraction pattern can be resolved with an integration time of around 2s. A comparatively slow growth rate of  0.03 layers per second was used, and the growth temperature was $\sim 570^\circ$C. 
\begin{figure}[ht]  \begin{center} \includegraphics [width=80mm]{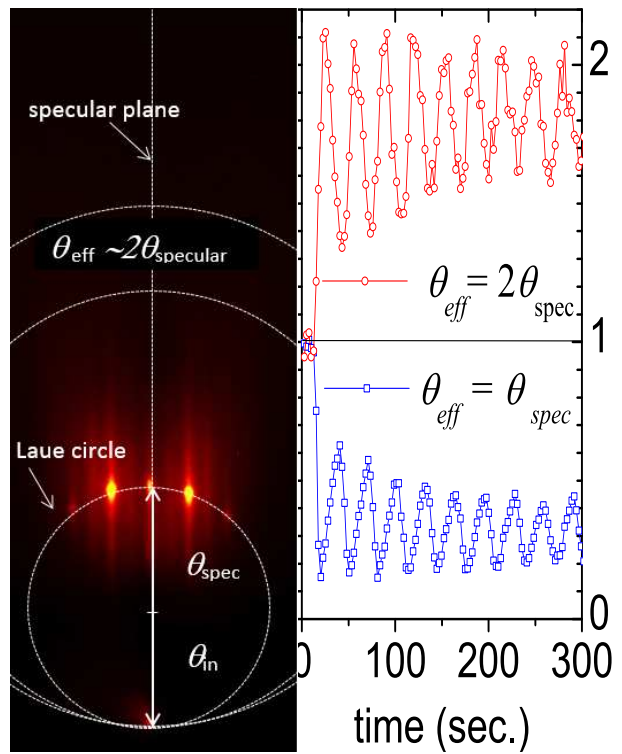}	\end{center}
		  		\caption{{Left: First (starting) image of a diffraction movie recorded during growth along the [110] direction with a 350 eV helium beam at $0.33^{\circ}$ incidence. Right: Time evolution of the scattered intensity $I(t)/I(t=0)$ around the specular angle $\theta_{spec}$ (blue line in the bottom) and at an effective angle twice larger (top red curve). As the intensity drops around the Laue circle, it increases at large scattering angle.}}
		  		\label{I_sur_I0}
\end{figure}
	Oscillations in the GIFAD scattered intensity during layer-by-layer growth are shown in Fig. \ref{I_sur_I0}. These have the same period and overall shape as RHEED oscillations carried out under the same conditions. However, unlike RHEED, due to the absence of penetration the phase of these oscillations is exactly the same whatever the diffraction order, whatever the angle of incidence or the crystal direction of the incident beam\cite{Atkinson2014}. 
	This lead us to associate the GIFAD oscillations to the variations of the surface reflectivity. When the helium beam impinges on a large flat terrace the reflectivity is close to $100\%$ but when obstacles are present along the helium trajectory, these atoms tends to be deflected to over specular scattering angles \cite{Pfandzelter}.  There is no penetration of the relatively low energy Helium atoms below the topmost layer therefore there are no interference effects which could alter the simple oscillations of the reflectivity. A maximum in the specular GIFAD scattered signal therefore will always correspond to the completion of a layer for true layer-by-layer growth. 
	
\begin{figure}[h]  \begin{center} \includegraphics [width=80mm]{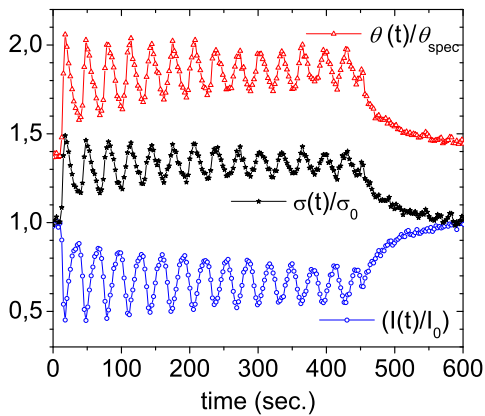}	\end{center}
		\caption{{The evolution in time of the first three moments of the distribution $I(\theta)$ are reported referred to their initial value (before growth) except for $\theta(t)$ for which the specular angle is a more natural reference.}}
		\label{mi}
\end{figure}

The detector records polar scattering angles up to $\sim 2.5^\circ$, which is a significant fraction of the scattering pattern. To some extend, it is therefore possible to track how the missing intensity on the Laue circle is distributed. This was already illustrated in specific 2D color plots in ref. \cite{Atkinson2014}. We present here a point of view in terms of an integrated scattering angle distribution $I(\theta_{eff})$ and the time evolution of its first three statistical moments; the integrated intensity $I(t)$, the mean value  $\bar{\theta(t)}$ and the width $\sigma(t)$. As illustrated on the left of Fig.\ref{I_sur_I0}, the effective scattering angle $\theta_{eff}$ associated with any point on the image is defined as the radius of the circle centered on the specular plane and intercepting the direct beam the points of interest. 
At the beginning of a new layer, the sudden drop in diffracted intensity is associated with a sudden apparition of scattering to super-specular angles. 
The figure \ref{I_sur_I0} shows that a factor two in effective scattering angle is enough to observe a complete reversal of the GIFAD oscillation. However, the intensity at half specular angle also tends to increase at the beginning of a layer, but overall the trend to larger angle dominates. This as illustrated in figure \ref{mi} by the increase of the mean scattering angle $\bar{\theta(t)}$. If both over and under specular intensities increase, then the width $\sigma(t)$ should be a reliable indicator of the growth. Indeed, the figure \ref{mi} shows that $\sigma(t)$ also oscillates with a nice triangular profile. Since, the evaluation of $\sigma(t)$ is not expected to depend significantly on whether or not diffraction is observed, $\sigma(t)$ it is probably a robust way for counting the layers when the identification of the Laue circle become problematic.

 The $I(t)$ curve in figure \ref{mi} corresponds to the intensity integrated on the whole detector. Oscillations are still present but the triangular shape has disappeared. In fact, if all scattered particle would reach the detector, no oscillation should be present anymore in $I(t)$! This indicates that, at the minimum of the intensity oscillation, around one half of the helium projectiles do not reach the detector, probably deflected to scattering angle beyond the restricted zone that was selected to reduce the size of the data storage. In summary, both the intensity restricted around the Laue circle as analyzed in \cite{Atkinson2014} and the width of the polar distribution are probably worth displaying to monitor of the growth.
	 
\begin{figure}[ht]  \begin{center} \includegraphics [width=70mm]{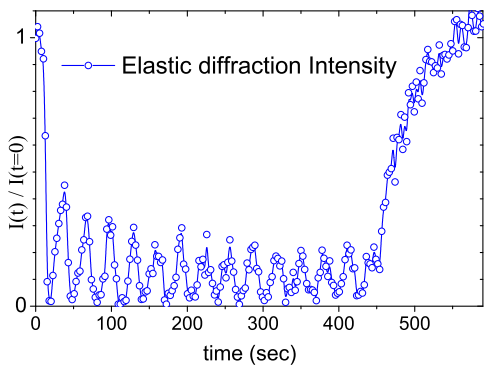}	\end{center}
		\caption{{Elastic diffraction intensity as a function of time. Even at the comparatively low growth rate, the elastic intensity drops down to negligible value at the beginning of each new layer suggesting this diffraction component requires a large surface coherence, i.e. that it is more sensitive to a low density defects.}}
		\label{elastic_intensity}
	\end{figure}

	We now focus on the evolution of the elastic diffraction. Its contribution, located on the Laue circle,is comparatively easy to extract \cite{Debiossac2014,Debiossac_V} by interpolating the background located below and above the Laue circle. The figure \ref{elastic_intensity}, shows that the intensity of this component drops to zero at the beginning of a new layer before slowly recovering. This suggests that elastic diffraction is more demanding in terms of surface coherence length, it is more sensitive to defect than inelastic diffraction which is observed to decrease but with a little contribution surviving all along the growth. 
	A simple interpretation could be that an atom can be elastically diffracted only if it probes a locally periodic surface, i.e. it does not encounter a defect. At variance, as long as the cumulated momentum spread induced by defect is smaller than a reciprocal lattice vector, the inelastic diffraction remains present. This could be the case if the projectile encounters defects located at few lattice units from its ``trajectory". 
	If the defect is closer producing a significant momentum transfer, the diffusion is likely to become completely incoherent and could possibly be modeled by classical scattering calculations. Detailed analysis of these three rich scattering contributions elastic, inelastic and diffuse may provide an insight into the different length-scales involved in island nucleation and coalescence during layer-by-layer growth.

\section{conclusion perspective}
We have demonstrated here that GIFAD is a technique which can be used to quantitatively test DFT models of surface reconstructions and the He-surface interaction potential for highly corrugated and complex surface reconstructions. Rocking curves together with a simple ray tracing analysis are also effective at rapidly confirming basic aspects of a surface reconstruction model. We have also shown that even for complex reconstructions,  there exist surface channeling directions with very weak corrugation which can therefore be modeled with a simple Fourier-like analysis - providing another means to check the surface reconstruction model. Finally we have shown that GIFAD is effective at monitoring layer-by-layer growth and that deconvolution of the intensity oscillations into components corresponding to elastic, inelastic and diffuse scattering may lead to time resolved  measurements of the mean distance between ad-atoms and provide new insights into island nucleation and coalescence.

	\section{Acknowledment}
	This work has been supported by Agence Nationale de la Recherche ANR-07-BLAN-0160 and ANR-2011-EMMA-003-01 and by the Triangle de la Physique
	(Contract No. 2012-040 T-GIFAD).

\end{document}